\documentclass[10pt,conference,letterpaper, twocolumn]{IEEEtran}

\usepackage[utf8]{inputenc}
\usepackage{graphicx}
\usepackage{stfloats}
\usepackage{amssymb}
\usepackage[cmex10]{amsmath}
\usepackage{color}
\usepackage{theorem}
\usepackage{array}
\usepackage{pifont}
\usepackage{psfrag}
\usepackage{hyperref}
\hypersetup{colorlinks=false,pdfborderstyle={/S/U/W 1},pdfborder=0 0 1}
\usepackage{cite}

\usepackage{bbm}

\usepackage{relsize}

\usepackage[ruled,vlined,titlenumbered]{algorithm2e}
\usepackage{tabularx}
\usepackage{arydshln}

\usepackage{tikz}
\usepackage{pgfplots}

\usepackage{booktabs}

\usetikzlibrary{chains,positioning,arrows,calc}

\usepgfplotslibrary{external}
\tikzsetexternalprefix{tikzfig/}
\pgfplotscreateplotcyclelist{mylist}{red,blue,black,yellow,brown}

\newcommand{\LCNOM}[1]{\dfrac{\sum\limits_{j \in \mathcal{Z}} \Big( a_j \prod_{\substack{\ell \in \mathcal{Z}\\ \ell \neq j}} (1-x \alpha^{#1} \beta^\ell ) \Big)}{ \prod\limits_{j \in \mathcal{Z}} (1-x \alpha^{#1} \beta^j)}}

\newcolumntype{Z}{>{\centering\arraybackslash}X}

\usepackage{ifthen}

\newcommand{\F}[1]{\ensuremath{\mathbb{F}_{#1}}}
\newcommand{\Fq}{\F{q}}
\newcommand{\Fx}[1]{\ensuremath{\F{#1}[x]}}
\newcommand{\Fxq}{\Fx{q}}

\newcommand{\refeq}[1]{(\ref{#1})}

\newtheorem{definition}{Definition}
\newtheorem{theorem}{Theorem}

\newtheorem{proposition}{Proposition}

\newtheorem{example}{Example}

\newcommand{\RS}[4]{\ensuremath{\mathcal{RS}(#1;#2,#3;#4)}}
\newcommand{\RSa}{\ensuremath{\mathcal{RS}}}
\newcommand{\VIRS}[1]{\ensuremath{\mathcal{VIRS}(n,k,s)}}

\newcommand{\mat}[2][\empty]{
  \ifthenelse{\equal{#1}{\empty}}
    {\ensuremath{\mathbf{#2}}}
    {\ensuremath{#2_{#1}}}
}

\newcommand{\vect}[2][\empty]{
  \ifthenelse{\equal{#1}{\empty}}
    {\ensuremath{\mathbf{#2}}}
    {\ensuremath{#2_{#1}}}
}

\DeclareMathOperator{\defi}{def}
\newcommand{\defeq}{\overset{\defi}{=}}

\newcommand{\defset}[1]{\ensuremath{D_\mathcal{#1}}}

\newcommand{\CYC}[4]{\ensuremath{\mathcal{C}(#1; #2,#3,#4)}}
\newcommand{\CYCa}{\ensuremath{\mathcal{C}}}
\newcommand{\LC}[4]{\ensuremath{\mathcal{L}(#1;#2,#3,#4)}}
\newcommand{\LCa}{\ensuremath{\mathcal{L}}}
\newcommand{\LCn}{\ensuremath{n_\mathcal{\ell}}}
\newcommand{\LCk}{\ensuremath{k_\mathcal{\ell}}}
\newcommand{\LCd}{\ensuremath{d_\mathcal{\ell}}}
\newcommand{\LCq}{\ensuremath{q_\mathcal{\ell}}}

\newcommand{\LCconst}{\ensuremath{e}}
\newcommand{\mulo}{\ensuremath{\mu}}

\newcommand{\HTconsta}{\ensuremath{b_1}}
\newcommand{\HTconstb}{\ensuremath{b_2}}
\newcommand{\muHT}{\ensuremath{d_0}}
\newcommand{\nuHT}{\ensuremath{\nu}}
\newcommand{\HTa}{\ensuremath{m_1}}
\newcommand{\HTb}{\ensuremath{m_2}}
\newcommand{\COS}[2]{\ensuremath{M_{#1}^{(#2)}}}
\newcommand{\NewBound}{\ensuremath{d^{\ast}}}

\begin{document}


\tikzset{external/system call={latex \tikzexternalcheckshellescape -halt-on-error -interaction=batchmode -jobname "\image" "\texsource"; dvips -o "\image".ps "\image".dvi}}

\title{Describing A Cyclic Code by Another Cyclic Code}

\IEEEoverridecommandlockouts

\author{\IEEEauthorblockN{Alexander Zeh \thanks{This work has been supported by DFG,
Germany, under grant BO~867/22-1.}}
\IEEEauthorblockA{Institute of Communications Engineering\\
University of Ulm, Ulm, Germany and \\
Research Center INRIA Saclay/\'Ecole Polytechnique, Paris, France\\
\texttt{alexander.zeh@uni-ulm.de}}
\and
\IEEEauthorblockN{Sergey Bezzateev}
\IEEEauthorblockA{Saint Petersburg State University\\ of Airspace Instrumentation\\
St. Petersburg, Russia\\
\texttt{bsv@aanet.ru}}
}

\maketitle

\begin{abstract}
A new approach to bound the minimum distance of $q$-ary cyclic codes is presented. The connection to the BCH and the Hartmann--Tzeng bound is formulated and it is shown that for several cases an improvement is achieved.

We associate a second cyclic code to the original one and bound its minimum distance in terms of parameters of the associated code.
\end{abstract}


\section{Introduction} \label{sec_intro}
Cyclic codes play an important role in coding theory and many communication systems. Their cyclic structure allows among other things efficient decoding methods.
For many cyclic codes, the minimum distance is not known, and hence we will investigate the minimum Hamming distance of $q$-ary cyclic codes in this contribution.
 
The Bose--Ray-Chaudhuri--Hocquenghem (BCH,~\cite{Hocquenghem_1959, Bose_RayChaudhuri_1960}) bound uses \textit{the longest} consecutive sequence in the defining set of the code to bound the minimum distance.
Its generalization, the Hartmann--Tzeng (HT,~\cite{Hartmann_GeneralizationsofBCHbound_1972,Hartmann_DecodingBeyondBCHBound_1974}) bound, is based on \textit{several} consecutive sets of zeros. Further generalizations are the contributions of Roos~\cite{Roos_GeneralBCHBound_1982, Roos_BoundforCyclicCodes_1983}, van Lint and Wilson~\cite{vanLint_OnTheMinimumDistance_1986},
Duursma and K\"otter~\cite{DuursmaKoetter_ErrorLocatingPairs_1994} and
Duursma and Pellikaan~\cite{Duursma_ASymmetricRoos_2006}.
Other approaches include the Boston bounds~\cite{Boston_CyclicCodesAlgebraicGeometry_2001} and the bound by Betti and Sala~\cite{BettiSala_NewLowerBound_2006}.

Our approach uses a second cyclic code --- the \textit{non-zero-locator code} --- to describe the defining set of the cyclic code which allows to bound its minimum distance.
It turns out that a good bound on the minimum distance is achieved, if the non-zero-locator code has low rate and a small distance.

This contribution is a generalization of our previous work~\cite{ZehWachterBezzateev_EfficientDecodingOfSomeClasses_2011,ZehWachterBezza_EfficientDecodingOfSomeClassesArxiv_2011}, where we used the power series expansion of a fraction of two co-prime polynomials and associated it with the code.
The advantage of this extension is that we can directly use well-known properties of cyclic codes to describe another cyclic code rather than abstract properties of power series expansions.
Further, this contribution is a generalization of~\cite{ZehWachterBezzateev_EfficientDecodingOfSomeClasses_2011,ZehWachterBezza_EfficientDecodingOfSomeClassesArxiv_2011} since the non-zero-locator code can be seen as a sum of several power series expansions.

Our contribution is structured as follows. We introduce necessary preliminaries of $q$-ary cyclic codes in Section~\ref{sec_preliminaries} and recall the HT bound.
Section~\ref{sec_loccode} gives the definition of the non-zero-locator code and proves the main theorem on the minimum distance. Single parity check and cyclic Reed--Solomon codes are used as non-zero-locator codes and the connection to the HT bound is shown in Section~\ref{sec_LocatorCodes}.
Section~\ref{sec_conclusion} concludes this contribution.


\section{Preliminaries} \label{sec_preliminaries}
Let $q$ be a power of a prime, let $\Fq$ denote the finite field of order $q$ and let $\Fxq$ denote the set of all univariate polynomials with coefficients in $\Fq$ and indeterminate $x$.
A $q$-ary cyclic code $\CYCa$ over $\Fq$ of length $n$, dimension $k$ and minimum distance $d$ is
denoted by \CYC{q}{n}{k}{d}. A codeword $c(x) = \sum_{i=0}^{n-1}c_i x^i$ of \CYC{q}{n}{k}{d} is a multiple of its generator polynomial $g(x) \in \Fxq$
with roots in \F{q^s}, where $n \mid (q^s-1)$.  Let $\alpha \in \F{q^s}$ be a primitive $n$th root of unity.
A cyclotomic coset $M_r$ is given by:
\begin{equation} \label{eq_cyclotomiccoset}
 M_r = \lbrace rq^j \bmod n, \; \forall j = 0,1,\dots,n_r-1 \rbrace,
\end{equation}
where $n_r$ is the smallest integer such that $rq^{n_r} \equiv
r \mod n$. It is well-known that the minimal polynomial $M_r(x) \in \Fxq$ of the element $\alpha^r$
is given by
\begin{equation} \label{eq_minpoly}
 M_r(x) = \prod_{i \in M_r} (x-\alpha^i).
\end{equation}
The defining set $D_{\mathcal{C}}$ of a $q$-ary cyclic code
\CYC{q}{n}{k}{d} is the set containing the indices of the zeros of the generator
polynomial $g(x)$ and can be partitioned into $m$ cyclotomic
cosets:
\begin{equation} \label{eq_definingset}
\begin{split}
D_{\mathcal{C}} & \defeq   \{i : \, g(\alpha^i)=0 \} = M_{r_1} \cup M_{r_2} \cup \dots \cup M_{r_{m}}.
\end{split}
\end{equation}
Hence, the generator polynomial $g(x) \in \Fxq$ of degree $n-k$ of \CYC{q}{n}{k}{d} is
\begin{equation}
 g(x) = \prod_{i=1}^{m} M_{r_i}(x).
\end{equation}
Let us recall a well-known bound on the minimum distance of cyclic codes.
\begin{theorem}[Hartmann--Tzeng (HT) Bound,~\cite{Hartmann_GeneralizationsofBCHbound_1972}] \label{theo_HT}
Let a $q$-ary cyclic code \CYC{q}{n}{k}{d} with the defining set
$D_{\mathcal{C}}$ be given.
Suppose there exist the integers $\HTconsta$, $\HTa$ and $\HTb$ with $\gcd(n,\HTa) = 1$ and $\gcd(n,\HTb) = 1$ such that
\begin{equation*} \label{eq_HTBound}
 \{ \HTconsta + i_1\HTa+i_2\HTb \mid 0 \leq i_1 \leq \muHT - 2,\; 0 \leq i_2 \leq \nuHT \} \subseteq \defset{\CYCa}.
\end{equation*}
Then, $d \geq \muHT+\nuHT$.
\end{theorem}
Note that for $\nuHT =0$ the HT bound becomes the BCH bound~\cite{Hocquenghem_1959, Bose_RayChaudhuri_1960}.
A further generalization was proposed by Roos~\cite{Roos_GeneralBCHBound_1982, Roos_BoundforCyclicCodes_1983} and van Lint and Wilson~\cite{vanLint_OnTheMinimumDistance_1986}.
Decoding up to the HT and the Roos bound was formulated by Feng and Tzeng~\cite[Section VI]{Feng-Tzeng:IEEE_IT1991}.

We consider cyclic Reed--Solomon (RS) codes~\cite{ReedSolomon_PolynomialCodesOverCertainFiniteFields_1960} for our approach and therefore recapitulate their definition in the following.
\begin{definition}[Cyclic Reed--Solomon Code] \label{def_CyclicRS}
Let $n$ be an integer dividing $q-1$ and let $\alpha$ denote an element of multiplicative order $n$ in \Fq{}.
Let $\delta$ be an integer.
Furthermore, let the generator polynomial $g_{\delta}(x) \in \Fxq$ be defined as:
\begin{equation*} \label{eq_GenPolyRS}
g_{\delta}(x) = \prod\limits_{i=\delta}^{\delta+n-k-1} (x-\alpha^i).
\end{equation*}
Then, a cyclic Reed--Solomon code over \Fq{} of length $n \mid q-1$ and dimension $k$, denoted by \RS{q}{n}{k}{\delta}, is defined by:
\begin{equation*} \label{eq_DefGenPolyRSCode}
\RS{q}{n}{k}{\delta} = \{ m(x) g_{\delta}(x) :  \deg m(x) < k \}.
\end{equation*}
\end{definition}
RS codes are maximum distance separable codes and their minimum distance $d$ is $d=n-k+1$.

\section{The Non-Zero-Locator Code} \label{sec_loccode}
We extend our earlier approach~\cite{ZehWachterBezzateev_EfficientDecodingOfSomeClasses_2011, ZehWachterBezza_EfficientDecodingOfSomeClassesArxiv_2011}, where we associated a
power series expansion of a fraction of two co-prime polynomials with the zeros of a cyclic code.
Now, we connect another cyclic code --- the so-called non-zero-locator code --- to a given cyclic code.

Let us establish a connection between the codewords of a cyclic code and the sum of power series expansions.
Let $c(x)$ be a codeword of a given $q$-ary cyclic code \CYC{q}{n}{k}{d} and let the set $\mathcal{Y}$ denote the set of indices of nonzero coefficients of $c(x)$
\begin{equation*}
c(x) = \sum\limits_{i \in \mathcal{Y}} c_{i} x^{i}.
\end{equation*}
Let $\alpha$ be an element of order $n$. Then, we have the following
relation for all $c(x) \in \CYC{q}{n}{k}{d}$:
\begin{equation} \label{eq_PowerSeriesCyclicCode}
 \sum\limits_{j=0}^{\infty} c(\alpha^j) x^j = \sum\limits_{i \in \mathcal{Y}} \frac{c_i}{1-x \alpha^i}.
\end{equation}
Now, we can define the non-zero-locator code.
\begin{definition}[Non-Zero-Locator Code] \label{def_locatorcode}
Let a $q$-ary cyclic code \CYC{q}{n}{k}{d} be given. Let $\alpha$ denote an $n$th root of unity.
Let $\gcd(n,\LCn) = 1$ and let $\beta$ be an $\LCn$th root of unity. Then, \LC{\LCq}{\LCn}{\LCk}{\LCd} is a non-zero-locator code of $\CYCa$ if there exists a $\mu \geq 0$ and an integer $\LCconst$, such that $\forall \, a(x) \in \LCa$ and $\forall \, c(x) \in \CYCa$:
\begin{equation*} \label{eq_locatorandcodeword}
\begin{split}
\sum_{j=0}^{\infty} c(\alpha^{j+\LCconst}) a(\beta^j) x^j \equiv 0 \bmod x^{\mulo-1},
\end{split}
\end{equation*}
holds.
\end{definition}
Before we prove the main theorem on the minimum distance of a cyclic code $\CYCa$, we describe Definition~\ref{def_locatorcode}.
We search the ``longest'' sequence
\begin{equation*}
c(\alpha^\LCconst) a(\beta^0), c(\alpha^{\LCconst+1}) a(\beta^1), \dots , c(\alpha^{\LCconst+\mulo-2}) a(\beta^{\mulo-2}),
\end{equation*}
that results in a zero-sequence of length $\mulo-1$, i.e., the product of the evaluated codeword $a(\beta^j)$ of the non-zero-locator code $\LCa$ and the evaluated codeword $c(\alpha^{j+\LCconst})$ of $\CYCa$ gives zero for all $j=0,\dots,\mulo-2$.

We require a root $\beta^j$ of the non-zero-locator code $\LCa$ at the position $j$ where the cyclic code $\CYCa$ has no zero.

We require $\gcd(n,\LCn) = 1$ to guarantee that
\begin{equation} \label{eq_coprime}
\begin{split}
 \gcd \Big( \prod_{j \in \mathcal{Y}}  (1-x\alpha^i\beta^j), & \prod_{j \in \mathcal{Y}} (1-x\alpha^m\beta^j) \Big) \\  & = 1 \, \forall i \neq m,
 \end{split}
\end{equation}
that we use for the degree calculation in the following. For the proof we refer to~\cite[Lemma 1]{ZehWachterBezzateev_EfficientDecodingOfSomeClasses_2011}.
We rewrite the expression of Definition~\ref{def_locatorcode} with~\refeq{eq_PowerSeriesCyclicCode} more explicitly. Let $\mathcal{Z}$ denote the set of indexes of nonzero coefficients of $a(x) \in \LCa$.
\begin{multline*} 
 \sum_{j=0}^{\infty}  c(\alpha^{j+\LCconst}) a(\beta^j) x^j = \sum\limits_{j=0}^{\infty} \sum\limits_{i \in \mathcal{Y}} c_i \alpha^{i(j+\LCconst)} a(\beta^j)x^j \\
 =  \sum\limits_{i \in \mathcal{Y}} c_i \alpha^{i\LCconst} \sum\limits_{j=0}^{\infty} \alpha^{ij} a(\beta^j)x^j
\end{multline*}
Using~\refeq{eq_PowerSeriesCyclicCode} for the two codewords $a(x)$ and $c(x)$ leads to:
\begin{multline*}
\sum\limits_{i \in \mathcal{Y}} c_i \alpha^{i\LCconst} \sum\limits_{j=0}^{\infty} \alpha^{ij} a(\beta^j)x^j =  \sum\limits_{i \in \mathcal{Y}} c_i \alpha^{i\LCconst} \sum\limits_{j \in \mathcal{Z}} \frac{a_j}{1-x\alpha^i \beta^j} \\
 = \sum\limits_{i \in \mathcal{Y}} c_i \alpha^{i\LCconst} \LCNOM{i},
\end{multline*}
and finally we obtain:
\begin{align}  \label{eq_locatorandcodeword2}
& \dfrac{\sum\limits_{i \in \mathcal{Y}} \Big( c_i \alpha^{i\LCconst} \sum\limits_{j \in \mathcal{Z}} \Big(a_j \prod\limits_{\substack{\ell \in \mathcal{Z}\\ \ell \neq j}} (1-x\alpha^i \beta^{\ell}) \Big) \prod\limits_{\substack{m \in \mathcal{Y}\\ m \neq i}} \prod\limits_{o \in \mathcal{Z}} (1-x\alpha^m \beta^o) \Big)  }{\prod\limits_{i \in \mathcal{Y}} \prod\limits_{j \in \mathcal{Z}} (1-x\alpha^i \beta^j)} \nonumber \\
& \equiv 0 \bmod x^{\mulo-1},
\end{align}
where the degree of the denominator is exactly $| \mathcal{Y} | \cdot | \mathcal{Z} |$ due to~\refeq{eq_coprime}. The degree of the numerator is smaller than or equal to $(|\mathcal{Y}|-1) \cdot | \mathcal{Z}| +  | \mathcal{Z}| -1$.
In the following we assume that the degree of the numerator is $(|\mathcal{Y}|-1) \cdot | \mathcal{Z}| +  | \mathcal{Z}| -1 = 
|\mathcal{Y}| \cdot |\mathcal{Z}| -1 $.

This leads to the following theorem on the minimum distance of a cyclic code $\CYCa$.
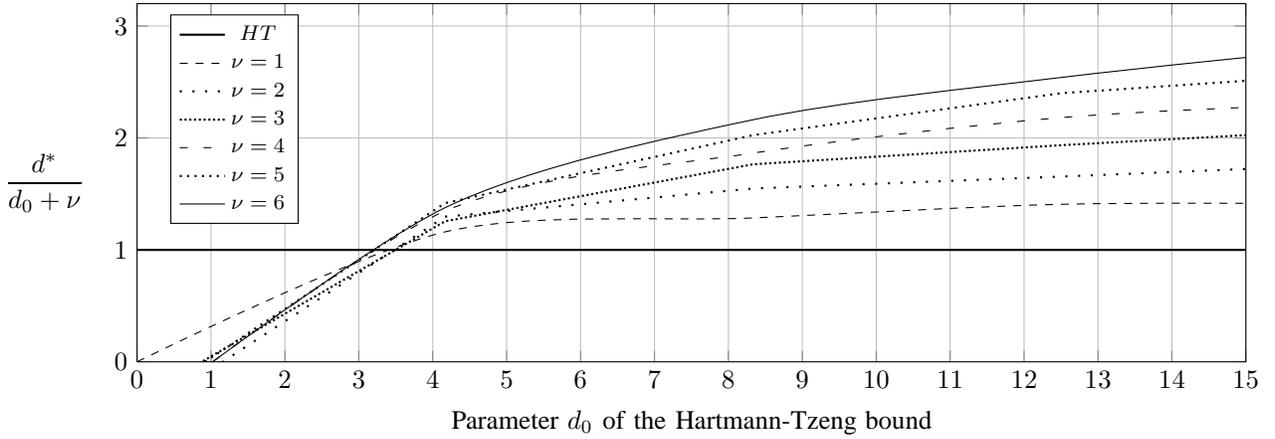
\begin{figure*}[t]
\normalsize
 \centering
 \begin{tikzpicture} [scale=1]
\begin{axis}
[	
	use fpu=false,
	legend style={font=\footnotesize},
	grid = both,
	ymin = 0,
	ymax = 3.2,
	xmin = 0,
	xmax= 15,
	xlabel=Parameter $\muHT$ of the Hartmann-Tzeng bound,
	ylabel style={rotate=270},
	ylabel = $\dfrac{\NewBound}{\muHT+\nuHT}$,
	width=0.9\textwidth,  
	height=0.35\textwidth,  
	cycle list name = mylist,
	legend pos= north west
]
	\addplot+[thick, black, domain=0:15] plot (\x,1) node[right] {$HT$};
	
	\addplot+[smooth,black,dashed, domain=0:100] plot (\x, {(ceil((1+2)*(\x-1)/2 +1))/(1+\x)});
	\addplot+[thick,black, loosely dotted,domain=0:100] plot (\x, {(ceil((2+2)*(\x-1)/2 +1))/(2+\x)});
	\addplot+[thick,black,densely dotted, domain=0:100] plot (\x, {(ceil((3+2)*(\x-1)/2 +1))/(3+\x)});
	\addplot+[smooth,black, loosely dashed, domain=0:100] plot (\x, {(ceil((4+2)*(\x-1)/2 +1))/(4+\x)});
	\addplot+[thick,black, dotted, domain=0:100] plot (\x, {(ceil((5+2)*(\x-1)/2 +1))/(5+\x)});
	\addplot+[smooth,black, domain=0:100] plot (\x, {(ceil((6+2)*(\x-1)/2 +1))/(6+\x)});
	\legend {$HT$,$\nuHT=1$, $\nuHT=2$,$\nuHT=3$, $\nuHT=4$,$\nuHT=5$, $\nuHT=6$}
\end{axis}
\end{tikzpicture}
\caption{Illustration of the fraction $\NewBound/(\muHT+\nuHT)$ of our bound $\NewBound$ of~\refeq{eq_dfracRS} to the Hartmann--Tzeng bound $\muHT+\nuHT$ for 
$\nuHT = 1,\dots,6$ and $\muHT=2,\dots,20$. The parameters of the HT bound are $\HTa=\nuHT+2$ and $\HTb = 1$ (see Table~\ref{tab_HT_ParityCheck}). We used a single parity check code as non-zero-locator code. Our bound $\NewBound$ is for $\muHT>3$ better than the HT bound. }
\label{fig_HTLocator1}
\end{figure*}
\begin{theorem}[Minimum Distance] \label{theo_minimumdistance}
Let a $q$-ary cyclic code \CYC{q}{n}{k}{d} with the associated non-zero-locator code \LC{\LCq}{\LCn}{\LCk}{\LCd} and the integers $\mu$ and $\LCconst$ be given with $\gcd(n,\LCn) = 1$, such
that~\refeq{eq_locatorandcodeword2} holds. Then, the minimum
distance $d$ of \CYC{q}{n}{k}{d} satisfies the following inequality:
\begin{equation} \label{inequality_for_d}
d \geq \NewBound \defeq \left \lceil \frac{\mulo}{\LCd} \right \rceil.
\end{equation}
\end{theorem}
\begin{IEEEproof}
For a codeword $c(x) \in \CYC{q}{n}{k}{d}$ of weight $d$ and codeword $a(x) \in \LC{\LCq}{\LCn}{\LCk}{\LCd}$ of weight $\LCd$, the degree of the denominator in~\refeq{eq_locatorandcodeword} is $d \cdot \LCd$. The numerator has degree at most $d \cdot \LCd-1$, and has to be greater than or equal to $\mulo-1$.
\end{IEEEproof}
\begin{example}[Binary Code of length $n=21$~\cite{Roos_BoundforCyclicCodes_1983, vanLint_OnTheMinimumDistance_1986}] Let the binary cyclic code \CYC{2}{21}{7}{8} with generator polynomial $g(x)$
\begin{equation*}
g(x) =  \COS{1}{21}(x) \cdot  \COS{3}{21}(x) \cdot \COS{7}{21}(x) \cdot \COS{9}{21}(x)
\end{equation*}
be given.

The defining set $\defset{\CYCa} = \COS{1}{21} \cup \COS{3}{21} \cup \COS{7}{21} \cup \COS{9}{21}$ of \CYC{2}{21}{7}{8} is:
\begin{equation*}
\begin{split}
\defset{\CYCa} = \{ 1,2,3,4,\square,6,7,8,9,\square,11,12,\square,14,15,16,\square,18 \},
 \end{split}
\end{equation*}
where the symbol $\square$ marks the index where $g(\alpha^i) \neq 0$.

We associate a single parity check code of length $\LCn = 5$, $\LCk=4$ distance $\LCd = 2$ as non-zero-locator code for \CYC{2}{21}{7}{8} according to Definition~\ref{def_locatorcode}.
For $\LCconst=0$ the subset of the defining set of \CYC{2}{21}{7}{8} and \LC{2^4}{5}{4}{2} is listed in Table~\ref{tab_Ex21RS}, where the product gives the a zero-sequence of length $13$.
The codewords $a(x) \in \LC{2^4}{5}{4}{2}$ ``fill'' the missing zeros of \CYC{2}{21}{7}{8} at position $0$, $5$ and $10$ in the interval $\left[0,12\right]$.
We have $\mu -1 = 13$ and therefore $\NewBound = \lceil (14)/2 \rceil = 7$.
\begin{table}[htbp]
\centering
\caption{Defining sets $\defset{\CYCa}$ and $\defset{\LCa}$ of the binary cyclic code \CYC{2}{21}{7}{8} and its non-zero-locator code \LC{2^4}{5}{4}{2} in the interval $\left[0,12\right]$.}
\label{tab_Ex21RS}
\begin{tabularx}{.95\columnwidth}{l|ZZZZZ;{2pt/2pt}ZZZZZ;{2pt/2pt}ZZZ}
$\defset{\CYCa}$ & $\square$ & 1 & 2 & 3 & 4 & $\square$ & 6 & 7 & 8 & 9 & $\square$ & 11 & 12 \\ 
$\defset{\LCa}$ & 0 & $\square$ & $\square$ & $\square$ & $\square$ & 0 & $\square$ & $\square$ & $\square$ & $\square$ & 0 & $\square$ & $\square$ \\ 
\end{tabularx}
\end{table}
The HT bound with parameters $\HTconsta=1$, $\HTa=5$, $\muHT=3$ and $\HTb=1$, $\nuHT=3$ gives also a lower bound of $7$ and the Roos bound gives $8$~\cite{vanLint_OnTheMinimumDistance_1986}, which is the minimum distance of $\CYC{2}{21}{7}{8}$.
\end{example}
The optimal non-zero-locator code \LCa{} for a cyclic code gives a zero sequence
\begin{equation*}
c(\alpha^{\LCconst}) a(\beta^0) , c(\alpha^{\LCconst+1}) a(\beta^{1}) , \dots , c(\alpha^{\LCconst+\mulo-2}) a(\beta^{\mulo-2})
\end{equation*}
of length $\mulo-1$ as in Definition~\ref{def_locatorcode}, such that $\NewBound$ of~\refeq{inequality_for_d} is maximized.

If we require a small cardinality of the defining set $\defset{\CYCa}$, the cardinality of the defining set $\defset{\LCa}$ of the non-zero-locator code should be large to obtain a long zero-sequence and therefore $\LCa$ should have a low rate $\LCk/\LCn$. 
On the other hand, the distance $\LCd$ of the non-zero-locator code \LCa{} should be small. 
\begin{table*}[!b]
\hrulefill
\centering
\caption{Defining sets $\defset{\CYCa}$ for $\HTconstb=1$ and $m$ of the HT bound~\refeq{eq_specialHT} and $\defset{\RSa}$ of the associated non-zero-locator code in the interval $\left[-(m-\nuHT)-1,m(\muHT-1)\right]$.}
\label{tab_HT_ReedSolomon}
\begin{tabularx}{2\columnwidth}{p{0.\columnwidth}|p{0.05\columnwidth}p{0.05\columnwidth}p{0.1\columnwidth}p{0.05\columnwidth}p{0.05\columnwidth}p{0.1\columnwidth};{2pt/2pt}p{0.01\columnwidth}p{0.05\columnwidth}p{0.1\columnwidth}p{0.03\columnwidth}p{0.05\columnwidth}p{0.1\columnwidth};{2pt/2pt}p{0.05\columnwidth}p{0.05\columnwidth}p{0.1\columnwidth}p{0.05\columnwidth}p{0.1\columnwidth}}
$\defset{\CYCa}$ & $\square$ & .. & $\square$ & 1 & .. & $\nuHT$+1 & $\square$ & .. & $\square$ & $m$+1 & .. & $m$+$\nuHT$+1 & $\square$ & .. & $\square$ & ..  & $\square$ \\ 
$\defset{\RSa}$ & 0 & .. & $m$-$\nuHT$-$2$ & $\square$ & .. & $\square$ & 0 & .. & $m$-$\nuHT$-$2$ & $\square$ & .. & $\square$ & 0 & .. &  $m$-$\nuHT$-$2$ & ..  &  $m$-$\nuHT$-$2$
\end{tabularx}
\end{table*}

\section{Beating the Hartmann--Tzeng Bound Using a Non-Zero-Locator Code} \label{sec_LocatorCodes}
\subsection{Normalization of HT Bound} \label{subsec_NormalizeHT}
Let us rewrite the HT bound as given in Theorem~\ref{theo_HT}. We multiply with the inverse of $\HTb$ modulo $n$.
Let \CYC{q}{n}{k}{d} be a $q$-ary cyclic code with the defining set $D_{\mathcal{C}}$.
Let
\begin{equation} \label{eq_specialHT}
\{ \HTconstb + i_{1}m+i_{2} : 0 \leq i_1 \leq \muHT-2, 0 \leq i_2 \leq \nu \} \subseteq D_{\mathcal{C}},
\end{equation}
where $\gcd(n,m)=1$. Then $d \geq \muHT + \nuHT$. 

Note that $m>\nuHT+1$. We refer to this representation of the HT bound in this section.
In the following, we consider a single parity check code as non-zero-locator code and outline the connection to a particular case of the HT bound.
The general case is then considered in Subsection~\ref{subsec_RS}, where cyclic RS codes are used as non-zero-locator codes.

\subsection{Parity Check Code as Non-Zero-Locator Code} \label{subsec_ParityCheck}
Let a $q$-ary cyclic code $\CYC{q}{n}{k}{d}$ with a subset of its defining set
with parameters $\muHT > 2$ and $\nuHT>0$ be given as stated in~\refeq{eq_specialHT}.
Furthermore, let $m= \nuHT+2$.

We associate a binary single parity check code as non-zero-locator code.
Let \LC{2}{\LCn}{\LCn-1}{2} be the cyclic non-zero-locator code with generator polynomial $g(x)=x-1$. We
assume $\gcd(n,\LCn) =1$ for the given cyclic code \CYC{q}{n}{k}{d}.
We illustrate the set of zeros of the cyclic non-zero-locator code \LCa{}, i.e., a single parity check code, for the cyclic code \CYC{q}{n}{k}{d} in Table~\ref{tab_HT_ParityCheck}.
A $\square$ represents the existence of a non-zero of the corresponding code \CYCa{} or \LCa{}. The sequence is illustrated in terms of parameters of the HT bound as in~\refeq{eq_specialHT}.
\begin{figure*}[t]
\normalsize
 \centering
\begin{tikzpicture} [scale=1]

\begin{axis}
[	
	use fpu=false,
	legend style={font=\footnotesize},
	grid = {both, ultra thin},
	ymin = 0,
	ymax = 3.2,
	xmin = 0,
	xmax= 15,
	xlabel=Parameter $\muHT$ of the Hartmann-Tzeng bound,
	ylabel style={rotate=270},
	ylabel = $\dfrac{\NewBound}{\muHT + \nuHT}$,
	width=0.9\textwidth,  
	height=0.35\textwidth,  
	cycle list name = mylist,
	legend pos= north west
]
	\addplot+[thick, black, domain=0:15] plot (\x,1) node[right] {$HT$};
	
	%
	%
	\addplot+[smooth,black,dashed, domain=0:100] plot (\x, {(ceil((6+2)*(\x-1)/2 +1))/(6+\x)});
	
	\addplot+[thick, black, loosely dotted,domain=0:100] plot (\x, {(ceil( ((6+3)*\x-6)/(3) )/(6+\x)});
	\addplot+[thick, black,densely dotted, domain=0:100] plot (\x, {(ceil( ((6+4)*\x-6)/(4) )/(6+\x)});
	\addplot+[smooth,black, loosely dashed, domain=0:100] plot (\x, {(ceil( ((6+5)*\x-6)/(5) )/(6+\x)});	
	\addplot+[thick, smooth,black, dotted, domain=0:100] plot (\x, {(ceil( ((6+6)*\x-6)/(6) )/(6+\x)});	
	
	\legend {$HT$,{$\LCd=2, \nu=6 $}, {$\LCd=3, \nu=6$}, {$\LCd=4, \nu=6$},{$\LCd=5,  \nu=6 $},{ $\LCd=6, \nu=6$}}
\end{axis}
\end{tikzpicture}
\caption{
Illustration of the fraction $\NewBound/(\muHT+\nuHT)$ of our bound $\NewBound$ of~\refeq{eq_dfracRS} to the Hartmann--Tzeng bound $\muHT+\nuHT$ for 
$\nuHT = 6$, $\muHT=2,\dots,20$, $\HTa = m$ and $\HTb=1$.
We used an RS code as non-zero-locator code with distance $\LCd = m-\nuHT$ (see Table~\ref{tab_HT_ReedSolomon}).
}
\label{fig_HTLocator2}
\end{figure*}
The considered code \CYCa{} has $\muHT-1$ sets of $\nuHT+1$ consecutive zeros, separated by one non-zero. The non-zero-locator code fills exactly this one non-zero.
\begin{table}[htbp]
\centering
\caption{Defining sets $\defset{\CYCa}$ for $\HTconstb=0$, $\HTa=m=\nuHT+2$, $\HTb=1$ and $\defset{\LCa}$ in the
interval $\left[-1,m(\muHT-1)-1\right]$.}
\label{tab_HT_ParityCheck}
\begin{tabularx}{\columnwidth}{p{0.025\columnwidth}|p{0.01\columnwidth}p{0.01\columnwidth}p{0.0005\columnwidth}p{0.06\columnwidth};{2pt/2pt}p{0.01\columnwidth}p{0.015\columnwidth}p{0.0005\columnwidth}p{0.08\columnwidth};{2pt/2pt}p{0.01\columnwidth}p{0.01\columnwidth}p{0.14\columnwidth};{2pt/2pt}p{0.01\columnwidth}}
$\defset{\CYCa}$ & $\square$ & 1 & .. & $m$-$1$ & $\square$ & $m$+$1$ & .. & 2$m$-1 & $\square$ &
.. & $m$($\muHT$-1)-1 & $\square$\\
$\defset{\LCa}$  & 0 & $\square$ & .. & $\square$ & 0 & $\square$ & .. & $\square$ & 0 & .. & $\square$ & 0 \\
\end{tabularx}
\end{table}

The parameters of the non-zero-locator code \LC{2}{\LCn}{\LCk}{\LCd} are:
\begin{equation*}
 \LCn = \nuHT+ 2, \quad  \LCk  = \nuHT +1, \quad  \LCd = 2
 \end{equation*}
and we have $\mulo - 1 = m \cdot (\muHT-1)+1$.
From \refeq{inequality_for_d} we obtain:
\begin{equation} \label{eq_boundsingeparitycheck}
 \begin{split}
 \NewBound & = \left \lceil \frac{m(\muHT-1)+2}{2} \right\rceil \\
 & = \left \lceil \frac{(\nuHT+2)\muHT - \nuHT}{2} \right\rceil.
 \end{split}
\end{equation}
In Fig.~\ref{fig_HTLocator1} we illustrate $\NewBound$ of~\refeq{eq_boundsingeparitycheck} for different parameters $\nuHT$ and $\muHT$ of the HT bound.
\begin{example}[Parity Code as Non-Zero-Locator Code]
Consider a cyclic code \CYC{q}{n}{k}{d} with the defining set $D_{\mathcal C}$
and let
\begin{equation*}
\{-5,-4,\square,-2,-1,\square,1,2,\square,4,5,\square \} \subseteq D_{\mathcal C}.
\end{equation*}
Furthermore let $\gcd(n,3)=1$. 
We associate a cyclic single parity check code of length $\LCn=3$ with \CYCa{} and illustrate
the corresponding zero-sequence in Table~\ref{tab_ex_ParityCheck}.
The zero-sequence has length $ \mulo - 1 = 13 $ and we obtain $\NewBound = \left \lceil (14)/2 \right\rceil = 7$.
\begin{table}[htbp]
\centering
\caption{Defining sets $\defset{\CYCa}$ for $\HTconstb=-5$, $\HTa=\nuHT+2=3$, $\HTb=1$ and $\defset{\LCa}$ in the
interval $\left[-6,6\right]$.}
\label{tab_ex_ParityCheck}
\begin{tabularx}{\columnwidth}{l|ccc;{2pt/2pt}ccc;{2pt/2pt}ZZZ;{2pt/2pt}ZZZ;{2pt/2pt}Z}
$\defset{\CYCa}$  & $\square$ & -5 & -4 & $\square$ & -2 & -1 & $\square$ & 1 & 2 & $\square$ & 4 & 5 & $\square$  \\
$\defset{\LCa}$ & 0 & $\square$ & $\square$ & 0 & $\square$ & $\square$ & 0 & $\square$ & $\square$ & 0 & $\square$ & $\square$ & 0 
\end{tabularx}
\end{table}
The HT bound gives for $\HTconstb=-5, m=3$ and $\muHT=5$, $\nuHT=1$ a lower bound of 
$d \geq 6$ on the minimum distance of $\CYCa$.
\end{example}

\subsection{Reed--Solomon Code as Non-Zero-Locator Code} \label{subsec_RS}
In the previous subsection we associated to $q$-ary cyclic code \CYCa{}, with a subset of its defining set with parameters $\HTa=m=\nuHT+2$ and $\HTb=1$ as stated in Theorem~\ref{theo_HT}, a single parity check code. Now we consider the case were $m > \nuHT+2$ and associate a RS code to the given $q$-ary cyclic code.

Let a $q$-ary cyclic code $\CYC{q}{n}{k}{d}$ with a subset of its defining set
with parameters $\muHT > 2$ and $\nuHT>0$ be given as stated in~\refeq{eq_specialHT}. Furthermore, let $m > \nuHT+2$.

In Table~\ref{tab_HT_ReedSolomon}, the HT bound~\refeq{eq_specialHT} with $i_1 = 0,\dots, \muHT-2$ and $i_2 =0,\dots,\nuHT$ is illustrated.
We choose as non-zero-locator code \LC{\LCq}{\LCn}{\LCk}{\LCd} a cyclic RS code with $\delta=0$ as in Definition~\ref{def_CyclicRS}. The parameters are:
\begin{equation*}
\LCn = m, \quad \LCk = \nuHT +1, \quad \LCd = m -\nuHT.
\end{equation*}
The $m-\nuHT-1$ consecutive zeros of the non-zero-locator code $\LCa{}$, i.e., a cyclic RS code of length $m$, fill the missing zeros of the given cyclic code $\CYC{q}{n}{k}{d}$.
We obtain for the ``zero''-sequence with length $\mulo  = m(\muHT-1) + m-\nuHT-1$.

Therefore, we obtain from~\refeq{inequality_for_d}:
\begin{multline} \label{eq_dfracRS}
\NewBound  = \left \lceil \frac{m(\muHT-1)+m-\nuHT}{m-\nuHT} \right\rceil \\
  = \left \lceil \frac{m\muHT-m+m-\nuHT}{m-\nuHT} \right\rceil \\
  = \left \lceil \frac{m\muHT-\nuHT}{m-\nuHT} \right\rceil.
\end{multline}
Note that for $m=\nuHT+2$ the non-zero-locator code is a single parity check code and we obtain
the result from~\refeq{eq_boundsingeparitycheck}.
Fig.~\ref{fig_HTLocator2} shows $\NewBound$ of~\refeq{eq_dfracRS} normalized to $\muHT+\nuHT$ for 
the same parameter $\nuHT=6$.
We varied the distance $\LCd$ of the non-zero-locator code.

Let us precise the cases where our bound $\NewBound$ is larger than the Hartmann--Tzeng bound $\muHT+\nuHT$.
\begin{proposition} \label{prop_OurBoundBetterThanHT}
Let a $q$-ary cyclic code \CYCa{} with a subset of its defining set with parameters $\muHT$, $\nuHT$, $\HTa=m$ and $\HTb=1$ as stated in Theorem~\ref{theo_HT} be given. Let $\LC{\LCq}{m}{\nuHT+1}{m-\nuHT}$ be the associated cyclic RS code as in Definition~\ref{eq_locatorandcodeword}.
Then, for 
\begin{equation*}
\muHT > m-\nuHT+1,
\end{equation*}
$\NewBound > \muHT + \nuHT$ holds.
\end{proposition}
\begin{IEEEproof}
From \refeq{eq_dfracRS} we have
\begin{multline*}
\NewBound = \left \lceil \frac{m\muHT-\nuHT}{m-\nuHT} \right\rceil \\
 = \left \lceil \frac{m\muHT-\muHT\nuHT +\muHT\nuHT - \nuHT}{m-\nuHT} \right\rceil \\
 = \left \lceil \muHT + \frac{(\muHT-1)\nuHT}{m-\nuHT} \right\rceil.
\end{multline*}
For $\NewBound > \muHT + \nuHT$, we need
\begin{equation} \label{eq_betterthanHT}
\begin{split}
 \frac{(\muHT-1)\nuHT}{m-\nuHT} & > \nuHT \\
 \muHT & > m-\nuHT+1  = \LCd+1
\end{split}
\end{equation}
\end{IEEEproof}
For $m-\nuHT = \LCd = 2$ the associated RS code is a single parity check code and our bound is better than the HT bound for $\muHT > 3$ (see Fig.~\ref{fig_HTLocator1}). 
Some other cases, where the distance of the associated RS code $m-\nuHT= \LCd $ is between two and six, are illustrated in Fig.~\ref{fig_HTLocator2}.

\section{Conclusion and Outlook} \label{sec_conclusion}
We presented and proved a new bound on the minimum distance of $q$-ary cyclic codes.
The used technique is based on a second cyclic code --- the so-called non-zero-locator code.
We used non-zero-locator codes that allow us to connect the Hartmann--Tzeng bound directly with our bound.
In detail, we used single parity check codes and RS codes and showed for which parameters our bound improves upon the HT bound.

Future work is the decoding up to our bound and the classification of cyclic codes, where the non-zero-locator code gives a good bound on the minimum distance.

\section*{Acknowledgement}
The authors wish to thank Antonia Wachter-Zeh and Daniel Augot for fruitful discussions.



\begin{thebibliography}{10}
\providecommand{\url}[1]{#1}
\csname url@samestyle\endcsname
\providecommand{\newblock}{\relax}
\providecommand{\bibinfo}[2]{#2}
\providecommand{\BIBentrySTDinterwordspacing}{\spaceskip=0pt\relax}
\providecommand{\BIBentryALTinterwordstretchfactor}{4}
\providecommand{\BIBentryALTinterwordspacing}{\spaceskip=\fontdimen2\font plus
\BIBentryALTinterwordstretchfactor\fontdimen3\font minus
  \fontdimen4\font\relax}
\providecommand{\BIBforeignlanguage}[2]{{%
\expandafter\ifx\csname l@#1\endcsname\relax
\typeout{** WARNING: IEEEtran.bst: No hyphenation pattern has been}%
\typeout{** loaded for the language `#1'. Using the pattern for}%
\typeout{** the default language instead.}%
\else
\language=\csname l@#1\endcsname
\fi
#2}}
\providecommand{\BIBdecl}{\relax}
\BIBdecl

\bibitem{Hocquenghem_1959}
A.~Hocquenghem, ``{Codes Correcteurs d'Erreurs},'' \emph{Chiffres (Paris)},
  vol.~2, pp. 147--156, Sep. 1959.

\bibitem{Bose_RayChaudhuri_1960}
\BIBentryALTinterwordspacing
R.~C. Bose and D.~K.~R. Chaudhuri, ``{On a class of error correcting binary
  group codes},'' \emph{Information and Control}, vol.~3, no.~1, pp. 68--79,
  Mar. 1960. [Online]. Available:
  \url{http://dx.doi.org/10.1016/S0019-9958(60)90287-4}
\BIBentrySTDinterwordspacing

\bibitem{Hartmann_GeneralizationsofBCHbound_1972}
\BIBentryALTinterwordspacing
C.~Hartmann and K.~Tzeng, ``{Generalizations of the BCH bound},''
  \emph{Information and Control}, vol.~20, no.~5, pp. 489--498, Jun. 1972.
  [Online]. Available: \url{http://dx.doi.org/10.1016/S0019-9958(72)90887-X}
\BIBentrySTDinterwordspacing

\bibitem{Hartmann_DecodingBeyondBCHBound_1974}
------, ``{Decoding beyond the BCH bound using multiple sets of syndrome
  sequences},'' \emph{Information Theory, IEEE Transactions on},
  vol.~20, no.~2, Mar. 1974.

\bibitem{Roos_GeneralBCHBound_1982}
\BIBentryALTinterwordspacing
C.~Roos, ``{A generalization of the BCH bound for cyclic codes, including the
  Hartmann-Tzeng bound},'' \emph{Journal of Combinatorial Theory, Series A},
  vol.~33, no.~2, pp. 229--232, Sep. 1982. [Online]. Available:
  \url{http://dx.doi.org/10.1016/0097-3165(82)90014-0}
\BIBentrySTDinterwordspacing

\bibitem{Roos_BoundforCyclicCodes_1983}
\BIBentryALTinterwordspacing
------, ``{A new lower bound for the minimum distance of a cyclic code},''
  \emph{IEEE Transactions on Information Theory}, vol.~29, no.~3, pp. 330--332,
  May 1983. [Online]. Available:
  \url{http://dx.doi.org/10.1109/TIT.1983.1056672}
\BIBentrySTDinterwordspacing

\bibitem{vanLint_OnTheMinimumDistance_1986}
\BIBentryALTinterwordspacing
J.~van Lint and R.~Wilson, ``{On the minimum distance of cyclic codes},''
  \emph{IEEE Transactions on Information Theory}, vol.~32, no.~1, pp. 23--40,
  Jan. 1986. [Online]. Available:
  \url{http://dx.doi.org/10.1109/TIT.1986.1057134}
\BIBentrySTDinterwordspacing

\bibitem{DuursmaKoetter_ErrorLocatingPairs_1994}
\BIBentryALTinterwordspacing
I.~M. Duursma and R.~Koetter, ``{Error-locating pairs for cyclic codes},''
  \emph{Information Theory, IEEE Transactions on}, vol.~40, no.~4, pp.
  1108--1121, Aug. 2002. [Online]. Available:
  \url{http://dx.doi.org/10.1109/18.335964}
\BIBentrySTDinterwordspacing

\bibitem{Duursma_ASymmetricRoos_2006}
\BIBentryALTinterwordspacing
I.~M. Duursma and R.~Pellikaan, ``{A symmetric Roos bound for linear codes},''
  \emph{J. Comb. Theory Ser. A}, vol. 113, pp. 1677--1688, Nov. 2006. [Online].
  Available: \url{http://portal.acm.org/citation.cfm?id=1226423}
\BIBentrySTDinterwordspacing

\bibitem{Boston_CyclicCodesAlgebraicGeometry_2001}
N.~Boston, ``{Bounding minimum distances of cyclic codes using algebraic
  geometry},'' \emph{Electronic Notes in Discrete Mathematics}, vol.~6, pp.
  385--394, 2001.

\bibitem{BettiSala_NewLowerBound_2006}
\BIBentryALTinterwordspacing
E.~Betti and M.~Sala, ``{A New Bound for the Minimum Distance of a Cyclic Code
  From Its Defining Set},'' \emph{Information Theory, IEEE Transactions on},
  vol.~52, no.~8, pp. 3700--3706, Jul. 2006. [Online]. Available:
  \url{http://dx.doi.org/10.1109/TIT.2006.876240}
\BIBentrySTDinterwordspacing

\bibitem{ZehWachterBezzateev_EfficientDecodingOfSomeClasses_2011}
A.~Zeh, A.~Wachter, and S.~Bezzateev, ``{Efficient Decoding of Some Classes of
  Binary Cyclic Codes Beyond the Hartmann-Tzeng Bound},'' in \emph{2011 IEEE
  International Symposium on Information Theory Proceedings (ISIT2011)}, St.
  Petersburg, Russia, Jul. 2011, pp. 1017--1021.

\bibitem{ZehWachterBezza_EfficientDecodingOfSomeClassesArxiv_2011}
------, ``{Decoding Cyclic Codes up to a New Bound on the Minimum Distance},''
  \emph{accepted for IEEE Transactions on Information Theory}, 2012.

\bibitem{Feng-Tzeng:IEEE_IT1991}
\BIBentryALTinterwordspacing
G.~L. Feng and K.~K. Tzeng, ``{Decoding cyclic and BCH codes up to actual
  minimum distance using nonrecurrent syndrome dependence relations},''
  \emph{IEEE Transactions on Information Theory}, vol.~37, no.~6, pp.
  1716--1723, 1991. [Online]. Available:
  \url{http://ieeexplore.ieee.org/xpls/abs\_all.jsp?arnumber=104340}
\BIBentrySTDinterwordspacing

\bibitem{ReedSolomon_PolynomialCodesOverCertainFiniteFields_1960}
\BIBentryALTinterwordspacing
I.~S. Reed and G.~Solomon, ``Polynomial {C}odes {O}ver {C}ertain {F}inite
  {F}ields,'' \emph{Journal of the Society for Industrial and Applied
  Mathematics}, vol.~8, no.~2, pp. 300--304, 1960. [Online]. Available:
  \url{http://dx.doi.org/10.1137/0108018}
\BIBentrySTDinterwordspacing

\end{thebibliography}
\end{document}